\documentstyle[12pt,epsf,epsfig]{article}

\def\ep{\epsilon}
\def\beq{\begin{equation}}
\def\eeq{\end{equation}}
\def\beqn{\begin{eqnarray}}
\def\eeqn{\end{eqnarray}}

\def\bea{\begin{eqnarray}}
\def\eea{\end{eqnarray}}

\def\GeV{{\;\rm GeV}}

\def\TeV{{\;\rm TeV}}

\def\cM{{\cal M}}
\def\cD{{\cal D}}
\def\cV{{\cal V}}
\def\epj#1#2#3{
        {\it Euro.\ Phys.\ J.\ }{\bf #1}, #2 (#3)}
\def\np#1#2#3{
        {\it Nucl.\ Phys.\ }{\bf #1}, #2 (#3)}
\def\pl#1#2#3{
        {\it Phys.\ Lett.\ }{\bf #1}, #2 (#3)}
\def\pr#1#2#3{
        {\it Phys.\ Rev.\ }{\bf #1}, #2 (#3)}

\def\ib#1#2#3{
        {\it ibid.\ }{\bf #1}, #2 (#3)}

\begin{document}

\begin{flushright}
\mbox{
\begin{tabular}{l} 
    FERMILAB-PUB-00/145-T
\end{tabular}}
\end{flushright}
\vskip 1.5cm
\begin{center}
\Large 
{\bf Radiative corrections to $Z b\bar{b}$ production} 

\vskip 0.7cm
\large
John Campbell and R.K. Ellis\\
\vskip 0.1cm
{\small Theory Department, Fermi National Accelerator Laboratory,
P.O. Box 500, Batavia, IL 60510} \\
June 27, 2000
\end{center}
\thispagestyle{empty}
\vskip 0.7cm

\begin{abstract}
We report on QCD radiative corrections to the process 
$p \bar{p}\rightarrow Z b \bar{b}$ 
in the approximation in which the $b$ quark is considered massless.
The implementation of this process in the general purpose Monte Carlo
program MCFM is discussed in some detail.
These results are used to investigate backgrounds to Higgs boson
production in the $ZH$ channel. We investigate the Higgs mass range
($100\GeV < m_H < 130\GeV$) for the Tevatron running at $\sqrt{s}=2 \TeV$.

\end{abstract}

\newpage

\section{Introduction}

In this paper we report on the calculation of the 
strong radiative corrections to the process
\beq
p + \bar{p} \rightarrow Z + b + \bar{b}\ .
\eeq
These results are obtained from a Monte Carlo program which allows
us to obtain predictions for any infra-red safe variable. 
Since the decays of the $Z$ are included we can perform cuts on the 
transverse momenta and rapidities of the final state leptons as well
as on the properties of the jets present in the event. This discussion
is similar to the calculation 
\beq
p + \bar{p} \rightarrow W + b + \bar{b}\ ,
\eeq
presented in ref.~\cite{EV}, although there are some 
differences. Ref.~\cite{EV} can
be seen as a first step towards the calculation of a vector boson plus
2 jets at ${\cal O}(\alpha_s^3)$, where the underlying partonic process
contains only 
quarks in the initial state. 
When considering the production of $Zb{\bar b}$ we
must also consider initial states containing two gluons, a further
stepping-stone towards a general $V+$~2~jet program.
With this in mind, we shall present our method in some detail,
with a general outline in Section 2 and further details presented
in the Appendix.

Much effort has been devoted to the study of Higgs production 
at the Tevatron running at $\sqrt{s}=2$~TeV~\cite{SUSYHIGGS}. 
These studies indicate that,
given enough luminosity, a light Higgs boson can be discovered at
the Tevatron  using the associated production channels $WH$ and $ZH$.
In Section 3 we perform an analysis in the $ZH$ channel
that incorporates as many of the backgrounds as
possible at next-to-leading order. Whilst
we use no detector simulation and do not attempt to include backgrounds
due to particle misidentification, 
the results presented here can provide a normalization
for more experimentally realistic studies. This is of importance since
more detailed studies are often performed using shower Monte Carlo
programs which can give misleading results for well separated jets. 

\section{Calculational overview}

In this section we will outline the implementation of the
matrix elements in our Monte Carlo program. We separate the discussion
according to the three types of contribution: the lowest order (Born)
processes, the virtual (loop) corrections and the real corrections
associated with additional soft or collinear radiation.

\subsection{Born processes}

\begin{figure}[ht]
\vspace{8.5cm}
\includegraphics{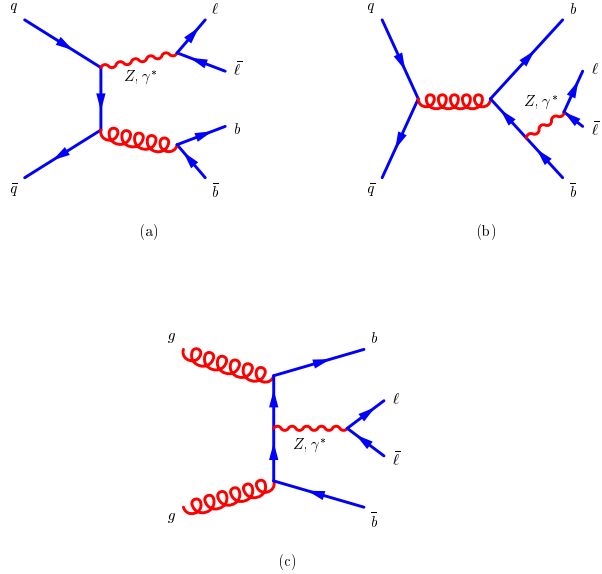}
\caption{Representative diagrams contributing to $Zb{\bar b}$
production at lowest order. A complete gauge-invariant set of diagrams
is included in the actual calculation.}
\label{zbbdiags}
\end{figure}

\begin{figure}[ht]
\vspace{10cm}
\includegraphics{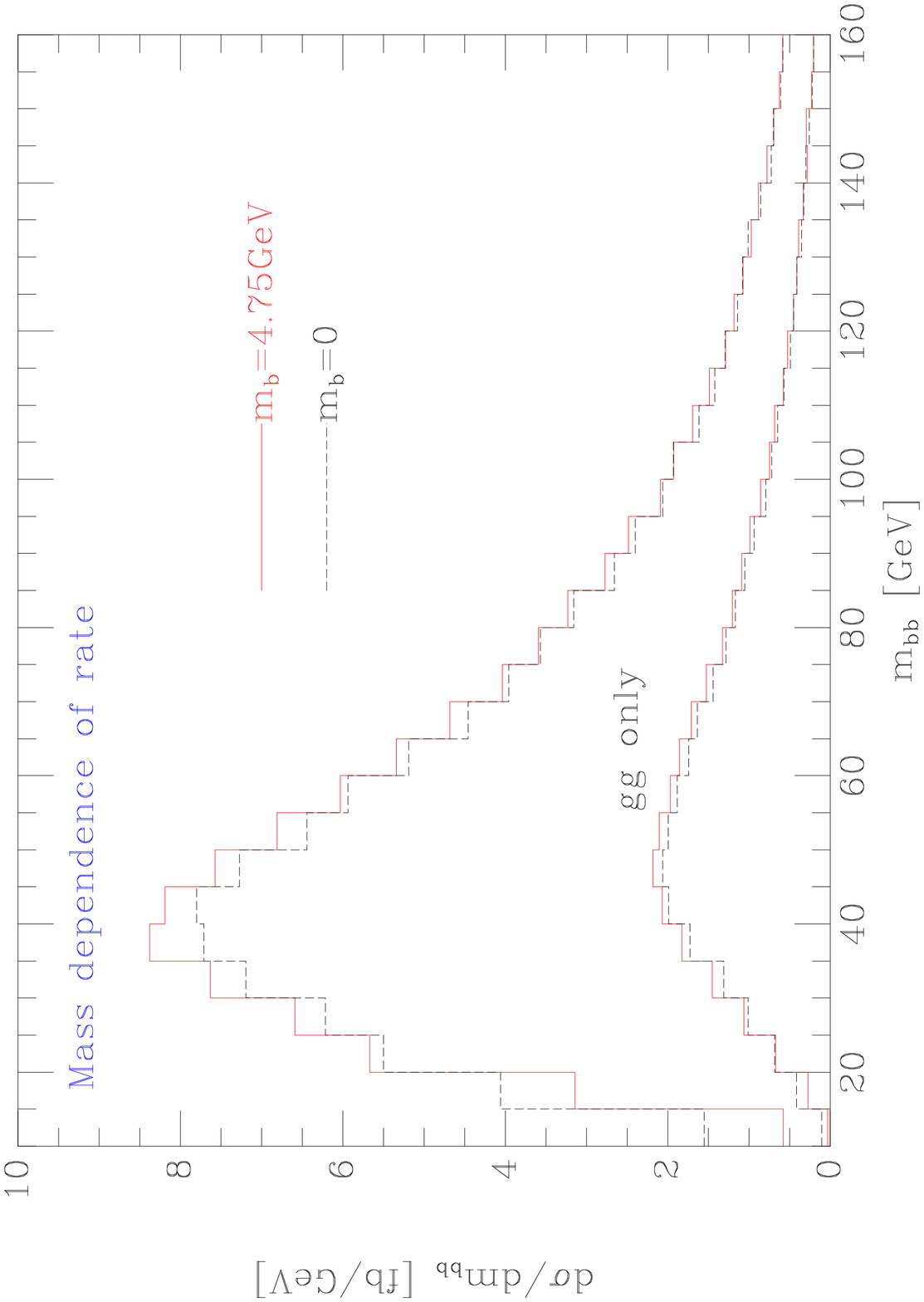}
\caption{Lowest order predictions showing the validity of the 
zero mass approximation}
\label{mdep}
\end{figure}

The Born processes which we are considering are 
\beqn \label{Born}
q + \bar{q} \rightarrow Z + b + \bar{b} \nonumber \\
g + g \rightarrow Z + b + \bar{b}
\eeqn
which are illustrated in Fig.~\ref{zbbdiags}.
For the $q\bar{q}$
case, the set of diagrams represented in (a) is the same as that
previously implemented for $W  b \bar{b}$ production in ref.~\cite{EV}.
The diagrams in (b) are related to those in (a) by crossing.
Diagrams (b) are absent in the $Wb{\bar b}$ case
since a $W$ does not directly couple
to a $b {\bar b}$ pair. For the same reason, the $gg$ process (c) is
also not present in the case of $W  b \bar{b}$ production and these
diagrams introduce a new set of matrix elements.
In all of the
matrix elements we will use the massless approximation for the
$b$-quarks. We expect this to be a good description for $b{\bar b}$
pairs produced with a large invariant mass, $m_{b{\bar b}}$. 
This expectation is borne out
by Fig.~\ref{mdep} where lowest order predictions with and without the 
$b$-quark mass are compared. As expected the corrections are of 
order $4 m_b^2/m_{b \bar{b}}^2$.  Note that Fig.~\ref{mdep} may
overestimate the mass effect since a fixed mass, rather than a running mass
which is smaller at high scale, is used. The basic jet cuts used in 
Fig.~\ref{mdep} to define the $b \bar{b}$ jets are,
\begin{eqnarray} \label{ur-cuts}
|y_{\mbox{jet}}| &<& 2.5\ , \nonumber \\
|p^T_{\mbox{jet}}| &>& 15\GeV \ ,\nonumber \\
\Delta R&>&0.7
\end{eqnarray}
where as usual
\begin{equation}
\Delta R =\sqrt{[\Delta \eta^2 + \Delta \phi^2]}\; .
\end{equation}

Fig.~\ref{mdep} also illlustrates the relative importance of the two 
sub-processes in Eq.~(\ref{Born}). The total cross-section and 
the contribution from $gg$-initiated processes are shown separately,
binned according to the $b{\bar b}$ invariant mass.
If we examine the total cross-sections (integrated over a large
range, $10\GeV < m_{b{\bar b}} < 160\GeV $), we find that the
gluon-gluon contribution is $36\%$ 
of the quark-antiquark result.
However, when we are later interested in examining this process as
a background to a $ZH(\to b{\bar b})$ signal, we will be interested in
the cross-sections over a limited range of $m_{b{\bar b}}$ close to a
Higgs mass $m_H$, where $100\GeV < m_H < 130\GeV$. We can see from the
figure that in this region the $gg$ contribution is a much more
significant fraction of the total, roughly $60\%$ of the $q{\bar q}$
cross-section in a window around $m_H=110\GeV$. The reason for the
importance of the $gg$ process in this region can be understood from
the contributing diagrams. By examination of Fig.~\ref{zbbdiags}(c) we see 
that, in contrast to the other diagrams in Fig.~\ref{zbbdiags},
$m_{b{\bar b}}$ may be large whilst
keeping all propagators on-shell.

\subsection{Real corrections}

The real matrix elements of Nagy and Trocsanyi (NT) given in  
ref.~\cite{NT} are implemented using a subtraction procedure~\cite{ERT}
which follows closely the treatment of Ref.~\cite{CS}. To illustrate
our method, we outline the subtraction terms that are needed for one
particular piece, namely the leading colour contribution to the $gg$
process. The actual numerical calculation that we have implemented
includes the full matrix elements, not just this leading colour piece.

We shall label the momenta of the partons as follows,
\beq
g(-p_1) + g(-p_2) \to Z(p_{34}) + b(p_5) + {\bar b}(p_6) + g(p_7),
\eeq
where all momenta are considered outgoing. In leading colour, it is
sufficient to consider one ordering of the gluons along the
fermion line, and to obtain the remaining terms by
permutation of the gluons~\cite{BerendsGiele}.
The real matrix elements are given by,
\beqn \label{M5}
|\cM_{\rm real}|^2
  & = &  8 e^4g^6 C_F N^3
 \times 
 \left( \sum_{\rm (6~perms)} 
\left| A(5_q^{h_5},1_g^{h_1},2_g^{h_2},7_g^{h_7},6_{\bar q}^{h_6})
 \right|^2 + {\cal O}\left(\frac{1}{N^2}\right) \right)
 \nonumber \\
 & \equiv & \sum_{\rm (6~perms)} |\cM_5(1_g,2_g,7_g) |^2
  + {\cal O}\left(C_F N\right),
\label{qqggg}
\eeqn
where we have used the (NT) notation for the
tree-level amplitude $A$ (see Eqs.~(NT:A43-A49)\footnote{We call the
reader's attention to the erratum of July, 2000 which corrects
some equations in ref.~\cite{NT}. The archive version has been
updated.}) and for simplicity we have suppressed all couplings of the
$Z$ boson and the sum over quark and lepton helicities.
As usual $C_F=4/3$ and $N=3$ is the number of colours.
The neglected terms are subleading in $N$.
Eq.~(\ref{M5}) is very similar to the leading-order term,
\beqn
|\cM_{\rm LO}|^2
 & = & 8 e^4g^4 C_F N^2
 \times 
 \left( \sum_{\rm (2~perms)} 
\left| m(5_q^{h_5},1_g^{h_1},2_g^{h_2},6_{\bar q}^{h_6})
 \right|^2 + {\cal O}\left(\frac{1}{N^2}\right) \right)
 \nonumber \\
 & \equiv & \sum_{\rm (2~perms)} |\cM_4(1_g,2_g) |^2
  + {\cal O}\left(C_F\right),
\label{qqgg}
\eeqn
which we will use to construct the subtraction terms.

\subsubsection{Dipole enumeration}

The matrix elements in Eq.~(\ref{qqggg}) become singular when
gluon 7 becomes soft and/or collinear with the quarks or either of the
other gluons. There are a number of methods for cancelling these
singularities at next-to-leading order~\cite{subtraction} and
we choose to use the dipole subtraction method of Catani and
Seymour~\cite{CS}. In this approach, the dipoles provide a convenient
and efficient way of enumerating the singularities and cancelling
them in a local fashion, including spin correlations.

A dipole consists of 3 partons, the first two of which combine to form
the emitter parton, $\widetilde{ai}$, and the third is the spectator $b$. 
The construction of the dipoles starts~\cite{ERT} 
from the eikonal factors present in the soft limit 
which are rewritten as,
\begin{equation} \label{eik}
\frac{2 p_a p_b }{p_a p_i \; p_b p_i} \equiv
\frac{2 p_a p_b }{p_a p_i+ p_b p_i} \Big [\frac{1}{p_a p_i} 
+ (a \leftrightarrow b)  \Big].
\end{equation}
The first term on the right hand side corresponds to the soft limit of the 
dipole with emitter $\widetilde{ai}$ and spectator $b$. 
The full dipole is given by 
the extension of Eq.~(\ref{eik}) to include collinear emission.
The dipole
must be labelled by the parton type of both emitter and emitted parton,
as well as whether the partons lie in the initial or final state.
For instance, the notation $\cD_{ii,\,gg}^{17,2}$
refers to a dipole where $\widetilde{17}$ is the initial-state gluonic
emitter splitting into gluon $1$ and parton $7$,
with respect to spectator parton $2$.

The dipole structure of the subtraction terms can be understood from the
soft gluon limit of the matrix elements~(\ref{qqggg}). In the soft
limit of a single colour ordering (permutation), gluon 7 can have
singularities with neighbouring partons only. Each of the six permutations
therefore gives rise to two dipoles, leading to twelve terms in all.
Hence we find,
\beq
|\cM|^2_{\rm dipole} = N
 \left[ 
          \cD_{ii,\,gg}^{17,2} + \cD_{ii,\,gg}^{27,1}
        + \cD_{if,\,gg}^{17,5} + \cD_{fi,\,qq}^{57,1} 
        + \cD_{if,\,gg}^{27,6} + \cD_{fi,\,qq}^{67,2} \right]
   \times \left| \cM_4(1_g,2_g) \right|^2 + (1 \leftrightarrow 2)
\label{real:dipoles}
\eeq
where the dipole functions are listed explicitly in
Appendix~\ref{app:dipoles}.

\subsection{Virtual corrections}

The virtual corrections to the processes in Eq.~(\ref{Born}) are taken
from the work of BDK, (Bern, Dixon and Kosower \cite{BDK}).
For other work on these parton subprocesses, see also ref.~\cite{BDKW}. 
In this paper the ultraviolet, infra-red and collinear singularities 
are controlled by continuing the dimension of space-time to 
$d=4-2\epsilon$ using the four-dimensional helicity scheme. 
The structure of the Monte Carlo program is such that virtual terms are
combined with the integrals of the dipole subtraction pieces defined above. 
When added together, $\epsilon$-poles cancel and a finite result is obtained.

To demonstrate this explicitly, we will examine the leading colour
contribution to the 1-loop diagrams. From BDK the leading 
$N$-contribution from the loop diagrams is
(cf. BDK, Eq.~(2.12)),
\beqn
 &&{\rm Re} \left( 2 \cM_{\rm LO} \cM^*_{\rm 1-loop} \right)
 = 8 e^4 g^4 C_F N^3 \left( \frac{\alpha_S}{2\pi} \right)
 (4\pi)^2 \nonumber \\
&& \times \Biggl[ 
 A_6^{\rm tree \, *}(5_q,1_g,2_g,6_{\bar q})
 A_{6;1}(5_q,1_g,2_g,6_{\bar q}) + {\cal O}
 \left(\frac{1}{N}\right) \Biggr]+ (1 \leftrightarrow 2).
\eeqn
Extracting only the pole pieces from the
leading term (see BDK, Eqs.~(8.5) and (8.7), together with the
renormalization in Eq.~(6.5)) yields,
\beqn
 && {\rm Re} \left( 2 \cM_{\rm LO} \cM_{\rm 1-loop}^* \right)^{\rm pole}
 = - \left( \frac{\alpha_S N}{2\pi} \right)
  |\cM(1_g,2_g)|^2 \times \Biggl[ \\
 && \frac{1}{\epsilon^2} \left(
  \left(\frac{s_{15}}{\mu^2}\right)^{-\epsilon}
 +\left(\frac{s_{12}}{\mu^2}\right)^{-\epsilon}
 +\left(\frac{s_{26}}{\mu^2}\right)^{-\epsilon} \right)
 +\frac{3}{2\epsilon}\left(\frac{s_{34}}{\mu^2}\right)^{-\epsilon}
 +\frac{1}{3\epsilon} \left( 11-\frac{2n_f}{N} \right)
 +\frac{7}{2}
 \Biggr]+ (1 \leftrightarrow 2), \nonumber 
\eeqn
where we have adopted the usual definition $s_{ij}=2 p_i \cdot p_j$.
We are now in a position to make a comparison with the dipole terms
that we presented in the previous section,
Eq.~(\ref{real:dipoles}). Inserting the appropriate integrals
(the functions $\cV^{end}$ from Appendix~\ref{app:dipoles}) we find
the counterterm from the real contribution to be,
\beqn
 |\cM|^2_{\rm counter}
 &=& \left( \frac{\alpha_S N}{2\pi} \right)
  |\cM(1_g,2_g)|^2 \times \frac{1}{2} \times \Biggl[ \nonumber \\
 && \frac{2}{\epsilon^2} \left(\frac{s_{12}}{\mu^2}\right)^{-\epsilon}  
 +\frac{1}{3\epsilon} \left( 11-\frac{2n_f}{N} \right)
 -\frac{\pi^2}{3} \nonumber \\
 &+& \frac{2}{\epsilon^2} \left(\frac{s_{15}}{\mu^2}\right)^{-\epsilon}
 +\frac{3}{2\epsilon}\left(\frac{s_{15}}{\mu^2}\right)^{-\epsilon}
 +\frac{1}{6\epsilon} \left( 11-\frac{2n_f}{N} \right)
 -\frac{\pi^2}{3} + 3 \nonumber \\
 &+& \frac{2}{\epsilon^2} \left(\frac{s_{26}}{\mu^2}\right)^{-\epsilon}
 +\frac{3}{2\epsilon}\left(\frac{s_{26}}{\mu^2}\right)^{-\epsilon}
 +\frac{1}{6\epsilon} \left( 11-\frac{2n_f}{N} \right)
 -\frac{\pi^2}{3} + 3
 \Biggr]+ (1 \leftrightarrow 2), \nonumber 
\eeqn
When we add these two contributions together, we see that the
$\epsilon$-poles cancel, leaving a finite term that is a combination
of logarithms and constants multiplying the lowest order squared matrix
elements,
\beqn
\lefteqn{
{\rm Re} \left( 2 \cM_{\rm LO} \cM^*_{\rm 1-loop} \right)^{\rm pole}
 + |\cM|^2_{\rm counter} = } && \nonumber \\
&&\left( \frac{\alpha_S N}{2\pi} \right) |\cM(1_g,2_g)|^2 \times \Biggl[
 \frac{3}{4}\ln\left(\frac{s_{34}}{s_{15}}\right)
 +\frac{3}{4}\ln\left(\frac{s_{34}}{s_{26}}\right)
 -\frac{\pi^2}{2}-\frac{1}{2} \Biggr] + (1 \leftrightarrow 2).
\eeqn
This result exemplifies the cancellation that occurs throughout our
calculation, both at sub-leading colour and in the other $q{\bar q}$
initiated sub-process (the details of which we have omitted here for
brevity). In general, when the poles present
in~\cite{BDK} are added to the integrated dipoles given in
Appendix~\ref{app:dipoles}, the result is extra finite terms
proportional to the lowest order matrix elements.

\section{Results}

\subsection{Basic features}

The standard model is specified by three gauge couplings; the top quark 
mass also leads to large corrections to tree graph results. For our purposes
we will replace these four parameters by $M_W,M_Z,\alpha(M_Z)$ and $G_F$, 
the values of which are given in Table~\ref{param1}.
\begin{table}[t]
\begin{center}
\begin{tabular}{|cc|cc|}
\hline
$M_Z,\Gamma_Z$ & $91.187,2.49\GeV$  &$ \alpha(M_Z)  $ &  1/128.89 \\
$M_W,\Gamma_W$ & $80.41,2.06\GeV$   &$ G_F  $& $  1.16639\times 10^{-5}$\\
\hline
\end{tabular}
\caption{Basic input parameters.}
\label{param1}
\end{center}
\end{table}
The coupling constants $e,g_W$ and $\sin^2 \theta_W$ 
are derived from these input parameters according to the definitions below,
\beqn
e^2 &=& 4 \pi \alpha(M_Z) \nonumber \\
g^2 _W &=& 4 \sqrt{2} G_F M_W^2 \nonumber \\
\sin^2 \theta_W &=&\frac{e^2}{g^2_W} \; .
\eeqn   
When defined in this way, these are effective parameters which
include the leading effects of top quark loops~\cite{Georgi}. 
In addition, we use the MRS98 parton distribution set~\cite{MRS}
with $\alpha_S(M_Z)=0.1175$.

\begin{figure}[ht]
\vspace{12.0cm}
\includegraphics{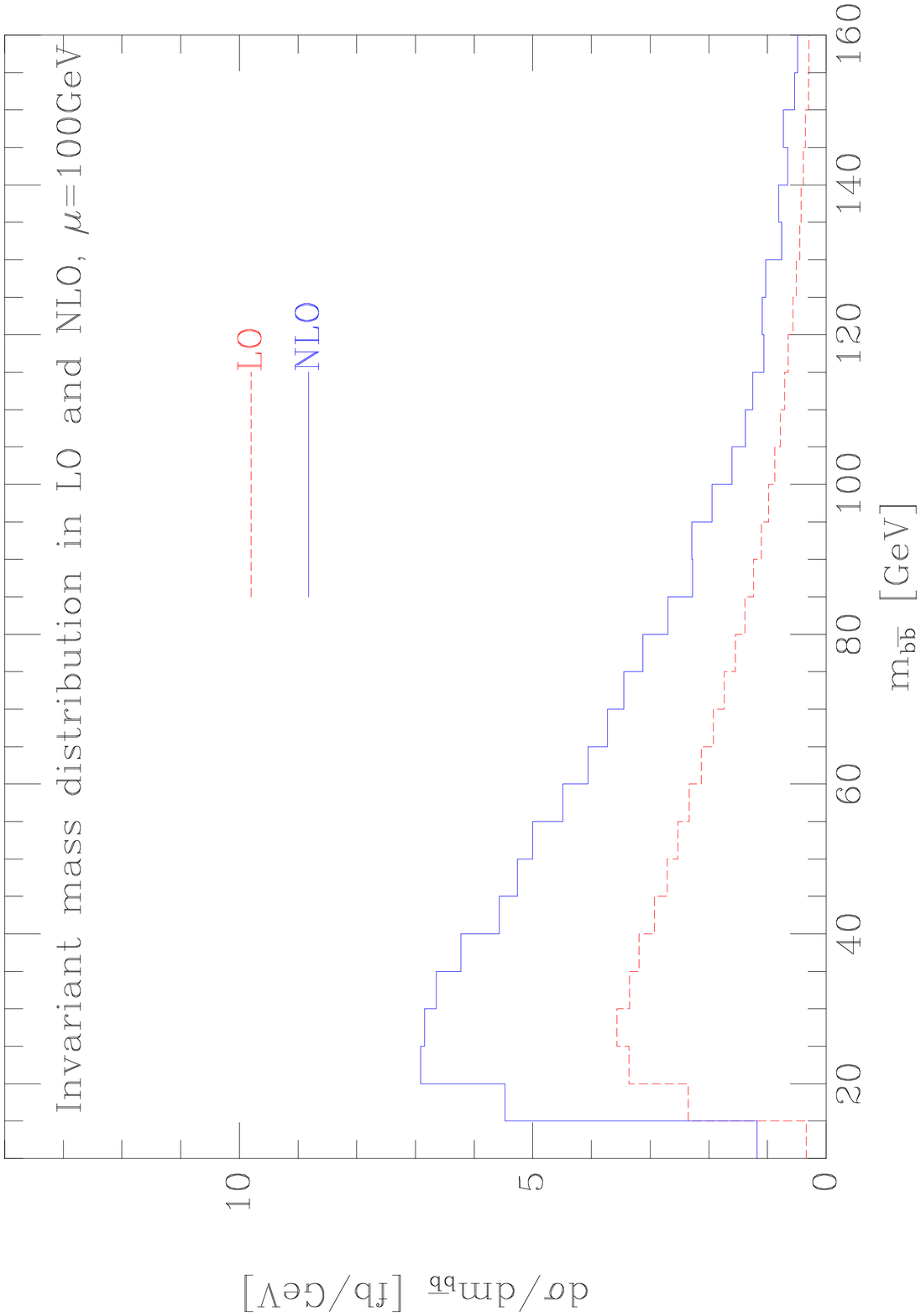}
\caption{The $b \bar{b} $ mass distribution in $b \bar{b}$ + missing energy 
events with the renormalization and factorization scale $\mu=100\GeV$.}
\label{Zbbar_m56_100}
\end{figure}

\begin{figure}[ht]
\vspace{10.0cm}
\includegraphics{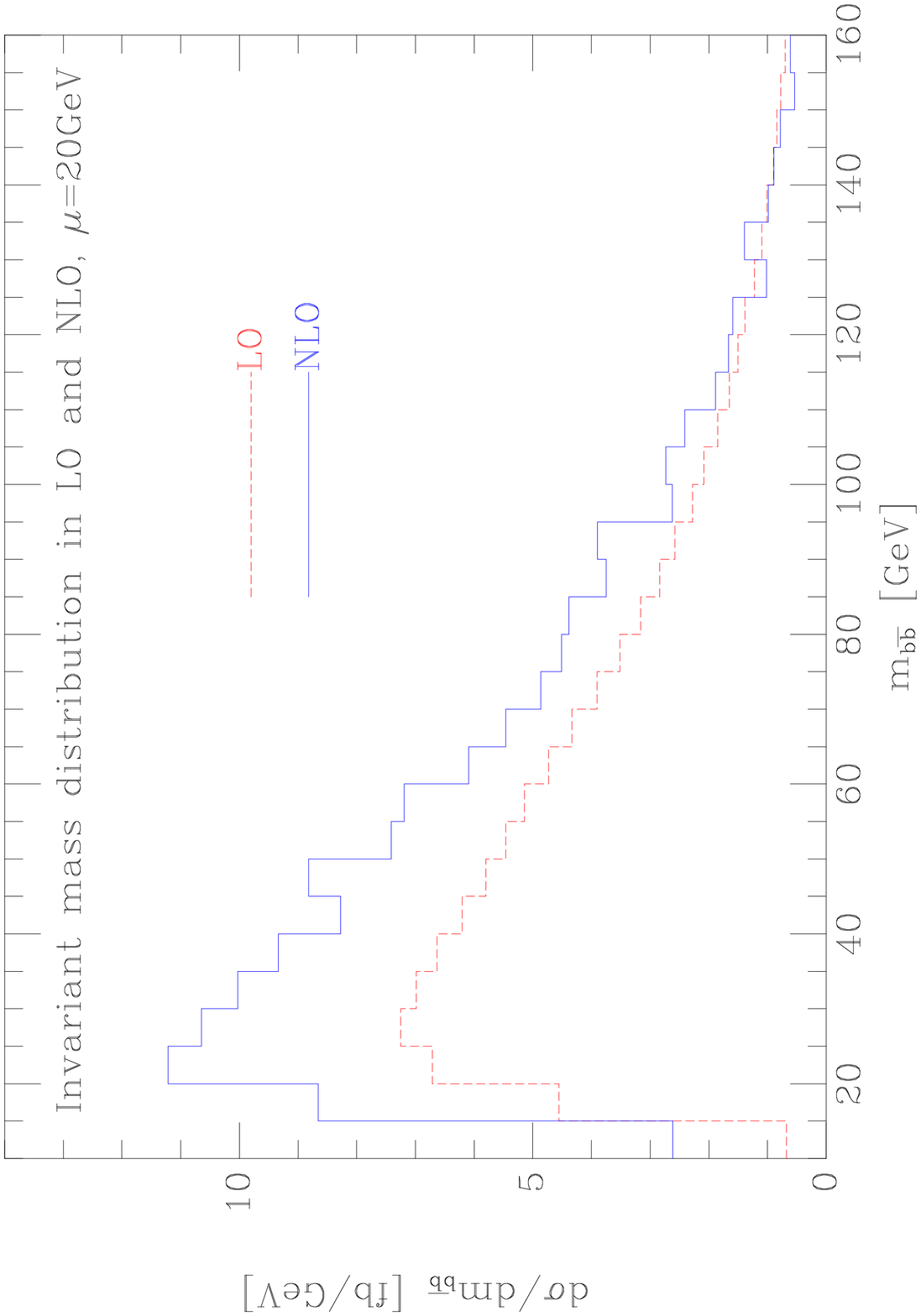}
\caption{The $b \bar{b} $ mass distribution in $b \bar{b}$ + missing energy 
events with the renormalization and factorization scale $\mu=20\GeV$.}
\label{Zbbar_m56_20}
\end{figure}

\def\Etmiss{\mathop{\not\!\! E_T}}
We apply our results to the phenomenological study of the 
$Zb{\bar b}$ background to $ZH$ production. 
We consider the channel $Z \rightarrow \nu \bar{\nu}$ which has 
a branching ratio of about $\sim 20\%$; the signature in this mode
is a $b \bar{b}$-pair and missing transverse energy, which we denote by 
$\Etmiss$. For convenience we impose the following cuts
on our Monte Carlo simulation,
and these will apply to all the results given in this section.  
We first require the observation of a 
$b$ and a $\bar{b}$ jet well separated from each other and from the 
direction in the transverse plane of the missing energy. The cuts we impose
(in addition to the basic jet cuts of Eq.~(\ref{ur-cuts})) are,
\begin{equation} \label{ZHcuts}
\begin{array}{rcl}
|y_b|, |y_{\bar{b}}| &<& 2\ , \\
|p^T_b|, |p^T_{\bar{b}}| &>& 15\GeV \ ,\\
\phi_{\; \Etmiss,b},\phi_{\; \Etmiss,\bar{b}},&>& 0.5, \\
\Delta R&>&0.7, \\
|\Etmiss| &>& 35\GeV. \\
\end{array}
\label{cutsznn}
\end{equation}
where $\phi_{\; \Etmiss,b}$ is the azimuthal angle between
the $b$-jet direction and the direction of the missing $E_T$. 
We also reject events which have additional jets in the
observed region
\beqn
|y_{\mbox{jet}}| &<& 2.5\ , \nonumber \\
|p^T_{\mbox{jet}}| &>& 15\GeV.
\label{cutszveto}
\eeqn 
We first examine the distribution of the cross-section
as a function of the $b\bar{b}$ mass. Our results are presented
at leading and next-to-leading order for a renormalization 
and factorization scale $\mu=100$~GeV and also for a much smaller
choice $\mu=20$~GeV in Figs.~\ref{Zbbar_m56_100}
and~\ref{Zbbar_m56_20}.
This smaller scale is chosen because, as
can be seen in Fig.~\ref{Zbbar_m56_20}, for this scale choice 
the distributions at LO and
NLO are comparable for $100< m_{b{\bar b}} < 160$~GeV.
We see that there is both a significant enhancement in
the total cross-section and a change in the shape
of the distribution when including
the radiative corrections. Moreover, for the lower choice of
scale, the total cross-section in the range 
$20< m_{b{\bar b}} < 160$~GeV is even larger and there is a
steepening of the distribution towards the peak at low $m_{b\bar{b}}$.

\begin{figure}[ht]
\vspace{10.0cm}
\includegraphics{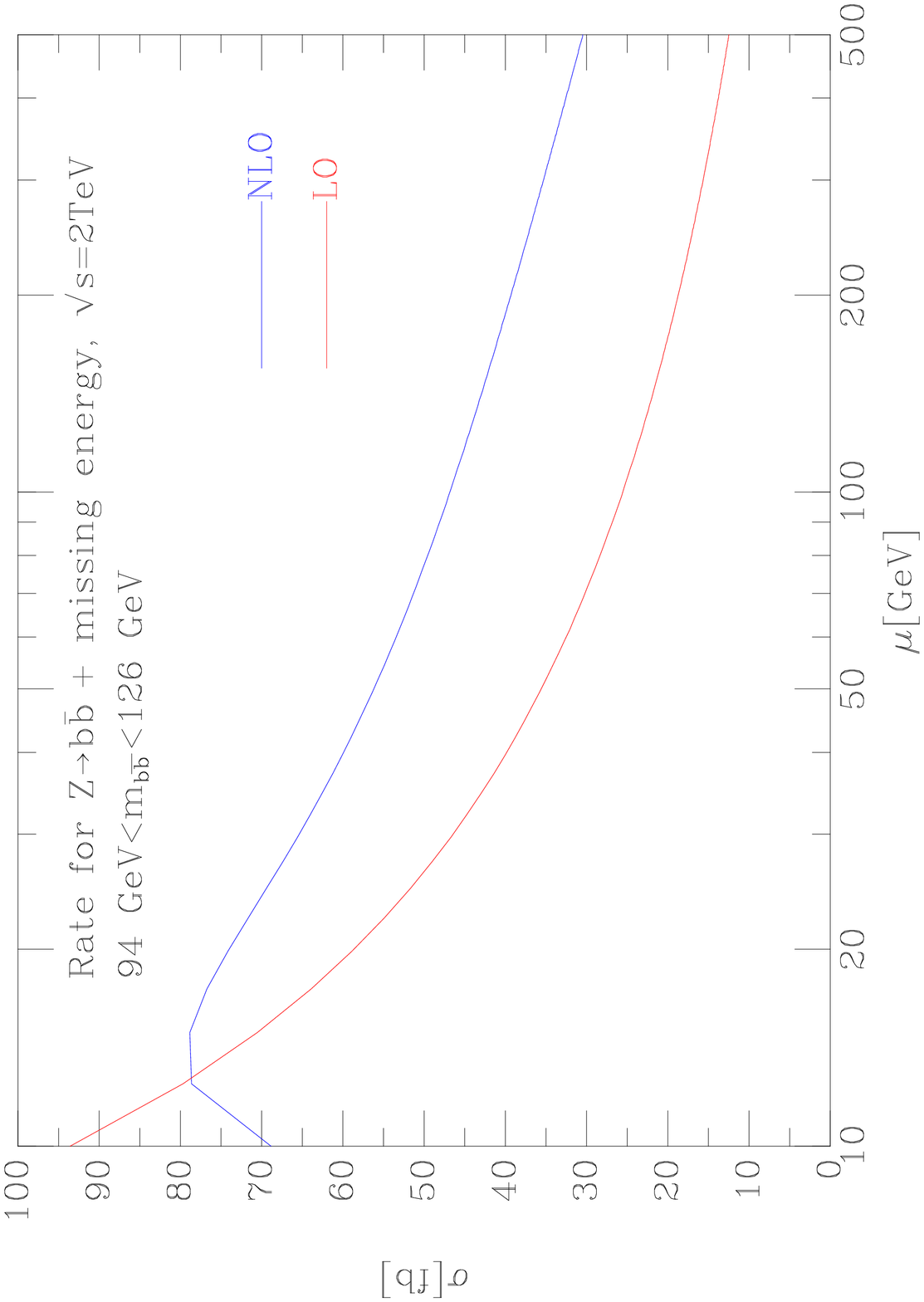}
\caption{The scale dependence of the $Zb{\bar b}$ result at both
leading and next-to-leading order, with $94<m_{b{\bar b}}<126$~GeV,
appropriate for $M_H =110$~GeV with $\Delta=11$~GeV.}
\label{zbb_mu}
\end{figure}

To further investigate the issue of scale dependence in this process,
we consider an integral of the cross-section over a restricted range
of $m_{b\bar{b}}$. We will consider such integrals in our analysis
of the Higgs search, where we shall integrate around the mass of
the putative Higgs boson. In particular, we will integrate over
\beq
m_H-\sqrt{2} \Delta < m_{b \bar{b}} < m_H + \sqrt{2} \Delta
\label{masswindow}
\eeq
where $\Delta$ is the mass resolution of the $b \bar{b}$-pair.
We assume that $\Delta=0.1 \times m_H$ and perform a gaussian
smearing with this resolution. The scale dependence of the
cross-section is shown in Fig.~\ref{zbb_mu}.
For low values of the renormalization and factorization
scale $\mu$ the NLO cross-section peaks; it is equal to the lowest
order result at $\mu \approx 15$~GeV.

\begin{figure}[ht]
\vspace{10.0cm}
\includegraphics{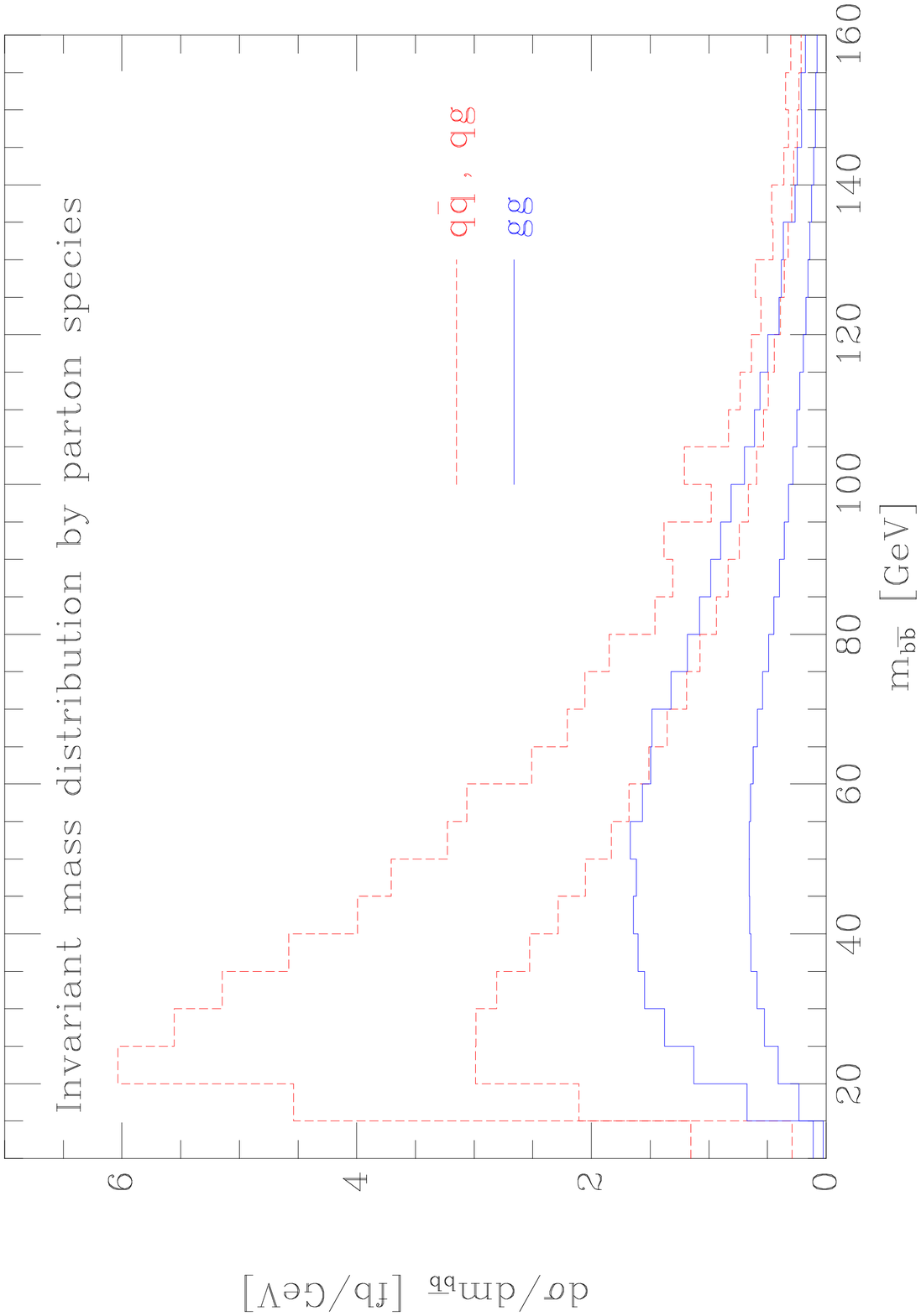}
\caption{The contribution of the different parton-parton sub-processes
to the cross-section, with the renormalization and factorization scale
$\mu = 100$~GeV. The lower curves represent the leading order results.}
\label{zbb_qq_gg}
\end{figure}

To conclude our discussion of the general features of the $Zb{\bar b}$
calculation at NLO, we assess the relative importance of the
different sub-processes. We have already seen that at leading order
the $gg$ process provides a significant contribution to the
cross-section, particularly for higher values of $m_{b{\bar b}}$. For
this reason, in Fig.~\ref{zbb_qq_gg} we show the
cross-section binned by $m_{b{\bar b}}$ and separated
into the contribution of the $gg$ component of parton luminosity 
and the contribution of the rest (beyond leading order we also have $qg$ 
initial states).

This figure shows that for the interesting range, $m_{b\bar{b}} \sim 100$~GeV,
the $gg$ contribution is even more important, relative to the
other sub-processes, than at leading order. In fact, if we divide the
next-to-leading order curve by the leading order one to obtain the
`$K$-factor' we find that for $m_{b \bar{b}}> 60$~GeV the ratio
$K_{q{\bar q}/qg} \approx 1.7$ whilst $K_{gg} \approx 2.5$.

\subsection{Higgs search}

Previous studies~\cite{SUSYHIGGS,Y,Mrenna} 
have shown that the channel $ZH$ where the 
$Z$ decays in neutrinos (observed in the detector as missing transverse energy)
and the Higgs boson decays to $b \bar{b}$ can give a significant signal for
the standard model Higgs boson with a mass below $130\GeV$.

The largest background to this signal is the $Zb{\bar b}$
process that we have discussed above. After this QCD background, the
largest other background at $2\TeV$ is from the diboson process
$Z^0 (\rightarrow \nu \bar\nu) Z(\rightarrow b \bar{b})$.
Significantly smaller backgrounds come from the processes
$W^{\pm *} (\to t(\rightarrow bW^+) \bar{b})$ and
$q^{\prime} t(\rightarrow bW^+) {\bar b}$, where in each case the
$W$ boson decays leptonically, but the charged lepton is not observed.
In addition contributions to the signal are present from 
$p \bar{p} \rightarrow WH$ where the $W$-decay lepton is missed;
correspondingly there are contributions to the background from  
$p \bar{p} \rightarrow W b \bar{b}$ and $p \bar{p} \rightarrow WZ$.

In our analysis, we will calculate the significances using our parton
level Monte Carlo MCFM in which the signal and largest backgrounds
are calculated beyond the leading order. In order to calculate the
signal we require the Higgs branching ratio into $b \bar{b} $ pairs,
which is a strong function of the Higgs mass.
The values of the branching ratio for the four values of the Higgs
mass that we will study, as well as additional parameters used
to calculate the other backgrounds, are shown in Table~\ref{param2}.
\begin{table}[t]
\begin{center}
\begin{tabular}{|cc|cc|}
\hline
$m_t,\Gamma_t$ & $175,1.4\GeV$      &$ \sin^2 \theta_W $  & 0.22853\\
$V_{ud}$&0.97500 & Higgs mass (br) & $100\GeV, (0.8119)$\\
$V_{us}$&0.22220 & Higgs mass (br) & $110\GeV, (0.7697)$\\
$V_{cd}$&0.22220 & Higgs mass (br) & $120\GeV, (0.6778)$\\
$V_{cs}$&0.97500 & Higgs mass (br) & $130\GeV, (0.5254)$\\
\hline
\end{tabular}
\caption{Additional parameters used for the signal and background
calculations for the Higgs search.}
\label{param2}
\end{center}
\end{table}
In order that our normalizations are clear, in Table~\ref{normalize}
we show the total cross-section for each process, together with the
results when appropriate branching ratios and lepton flavour sums are
included. The values shown for the
$W b \bar{b}$ and $Z b \bar{b}$ processes only have the basic cuts,
Eq.~(\ref{ur-cuts}); for $t \bar{t}, WH$ and $ZH$ the rates shown
are the total cross sections with no cuts.
\begin{table}[t]
\begin{center}
\begin{tabular}{|c|c|c|c|}
\hline
Process & $\sigma$ [fb] & With BR & $\sigma_{BR}$ [fb] \\
\hline
$WH$, $m_H=100$ & $274$
&$W^+ (\to e^+ \nu) H (\to b {\bar b}) \times 4$ & 48.9 \\
$ZH$, $m_H=100$ & $162$ 
&$Z (\to \nu_e {\bar \nu}_e) H (\to b {\bar b}) \times 3$ & 25.3  \\
$Z b \bar{b}$, $60< m_{b \bar{b}}<160$ & 5290
&$Z (\to \nu_e {\bar \nu}_e) b \bar{b} \times 3$ & 1010 \\
$ZZ$  & 1240 
&$Z (\to \nu_e {\bar \nu}_e) Z (\to b {\bar b}) \times 3 \times 2$ &71.4 \\
$W b \bar{b}$, $60< m_{b \bar{b}}<160$ & 8060
&$W^+ (\to e^+ \nu) b \bar{b} \times 4$ & 1770 \\
$WZ$ & 3600
&$W^+ (\to e^+ \nu) Z (\to b {\bar b}) \times 4$ & 119\\
$W^{\pm *} (\to t(\rightarrow bW^+) \bar{b})$ & 449
&$W^{\pm *} (\to t (\to b W^+ (\to e^+ \nu)) {\bar b})\times 4$ & 98.8\\
$q^{\prime} t(\rightarrow bW^+) {\bar b}$ & 81
&$q^{\prime} t (\to b W^+ (\to e^+ \nu)) {\bar b} \times 4$ & 35.6 \\
\hline
\end{tabular}
\caption{Values of the total cross-sections for normalization purposes} 
\label{normalize}
\end{center}
\end{table}

For each choice of the Higgs boson mass, the signal and backgrounds
are integrated over an $m_{b\bar{b}}$ masss range as
in~Eq.~(\ref{masswindow}). For the backgrounds we also reject events
with additional observed leptons,
\beqn
|y_l| &<& 2, \nonumber \\
|p^T_l| &>& 10\GeV .
\eeqn

Our results at $2\TeV$ are given in Table~\ref{table_z_2},
where we have used a double $b$-tagging
efficiency $\epsilon_{b\bar{b}}=0.45$.
\begin{table}
\begin{center}
\begin{tabular}{|l|c|c|c|c|c|}
\hline
$m_H$~[GeV] & 100  &110  & 120 & 130     \\
\hline
$Z^0 (\rightarrow \nu \bar\nu) H(\rightarrow b \bar{b})$
 & $5.8$ & $4.4$ & $2.9$ & $1.8$   \\
$W (\rightarrow \ell \nu) H(\rightarrow b \bar{b})$
 &  $0.7$ & $0.5$ & $0.3$ & $0.2$   \\
Total $S$ & $6.5$ & $4.9$ & $3.2$ & $2.0$   \\
\hline
\hline
$Z^0 (\rightarrow \nu \bar\nu) g^*(\rightarrow b \bar{b})$
 & $23.3$ & $21.0$ & $17.5$ & $15.8$   \\
$Z^0 (\rightarrow \nu \bar\nu) Z(\rightarrow b \bar{b})$
 & $10.8$ & $6.9$ & $3.3$ & $1.4$  \\
$W^\pm (\rightarrow \ell \bar\nu) g^*(\rightarrow b \bar{b})$
 & $2.7$ & $2.4$ & $1.6$ & $1.1$  \\
$W^\pm (\rightarrow \ell \bar\nu) Z(\rightarrow b \bar{b})$
 & $1.4$ & $0.8$ & $0.4$ & $0.2$  \\
$W^{\pm *} (\to t(\rightarrow bW^+) \bar{b})$&
 $0.6$ & $0.6$ & $0.6$ & $0.6$   \\
$q^{\prime} t(\rightarrow bW^+) {\bar b}$ & 
 $0.2$ & $0.2$ & $0.2$ & $0.2$  \\
Total $B$ & $39.0$ & $31.9$ & $23.6$ & $19.3$   \\
\hline
$S/B$ & $0.17$ & $0.15$ & $0.14$ & $0.11$ \\
\hline
$S/\sqrt{B}$ & $1.04$ & $0.87$ & $0.66$ & $0.46$ \\
\hline
\end{tabular}
\caption{Signal, backgrounds and significance for the $Z$-channel
at $\sqrt{s}=2~\TeV$}
\label{table_z_2}
\end{center}
\end{table}
When we compare these results to those of the SUSY-Higgs
study~\cite{SUSYHIGGS} we immediately see a difference in some
of the channels. In particular, by including the radiative corrections
to the $Zb{\bar b}$ process we have increased this background
considerably. Furthermore, the contribution to the signal
from the $WH$ process and all the backgrounds involving a $W$ are
significantly lower than in~\cite{SUSYHIGGS}. The reason for this
discrepancy is clear. In our Monte Carlo, all these channels involve
an unobserved lepton, which must therefore have a very low $p_T$ or
be at high rapidity. With a detector simulation, such as is used in the
SUSY-Higgs study, these unobserved leptons may be central and thus
produce large contributions to the cross-sections, whilst being
unobserved by virtue of misidentification.

\begin{figure}[ht]
\vspace{10.0cm}
\includegraphics{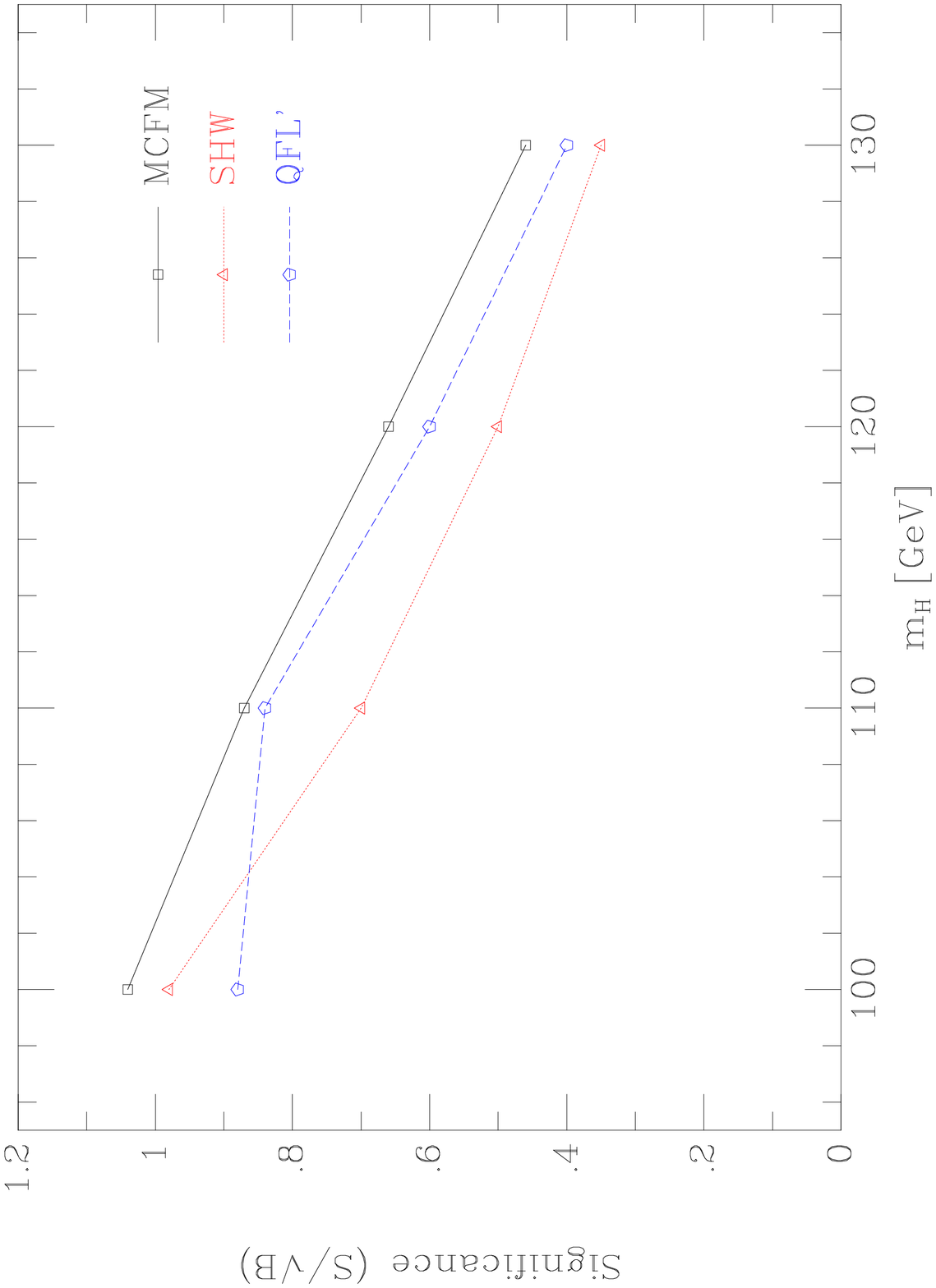}
\caption{The significance obtained from our Monte Carlo MCFM, as
presented in Table~\ref{table_z_2}, compared to the results obtained
by the SUSY-Higgs workshop~\cite{SUSYHIGGS}. Note that the MCFM significance
should not be considered realistic because of the lack of detector effects.}
\label{signif}
\end{figure}

To illustrate the effect of these differences on the significance
of the different analyses, in Fig.~\ref{signif} we compare our
MCFM results with two approaches from ref.~\cite{SUSYHIGGS}.
The apparent similarity of the significances plotted in Fig.~\ref{signif}
is the result of two effects. We believe that the significance of
ref.~\cite{SUSYHIGGS} is too high because, the 
$Z b \bar{b}$  and $ZZ$ backgrounds are underestimated 
in the light of the work presented here and in Ref.~\cite{CE}.
On the other hand, our significances are certainly too high, because
due to the lack of a full detector simulation, the backgounds 
from processes with mis-identified leptons are underestimated.  

\section{Conclusions}
We have presented the first results on the strong radiative corrections
to the $Z b \bar{b}$ process. The results indicate large radiative 
corrections which can significantly change the estimates of the 
backgrounds to the process $p \bar{p} \rightarrow ZH$ at the Tevatron.
We find that the significance $S/\sqrt{B}$ could be  
less than the result in the report of the 
SUSY-Higgs working group. 
The results described in this paper do not represent a full analysis
of the potential to see the Higgs signature in the $ZH$ channel at the 
Tevatron. The most significant shortcoming is the lack of a full detector 
simulation. 

The primary change is the increase in the size of the $Z b \bar{b}$-background
because of the inclusion of NLO effects. Our calculation underscores 
the importance of having an experimental determination of the $Z b \bar{b}$
background, either by relating it to observed  $Z b \bar{b}$ events at 
lower $m_{b \bar{b}}$ or by relating it to $Z$+two jet events.

\begin{center}
{\bf Acknowledgements}
\end{center}
This work was supported in part by the U.S. Department of Energy
under Contract No. DE-AC02-76CH03000.

\appendix
\section{Dipole formulae}
\label{app:dipoles}

This appendix details the exact dipole functions that we have used in
the subtraction of the real integral, as well as the integrals of these
dipoles that must be added to the loop contributions. We note that
these functions differ slightly from those given in~\cite{CS}
since they carry no colour factors. In addition, these are the dipoles
appropriate for the four-dimensional helicity scheme, which is the scheme
of choice for the calculation of complicated virtual corrections.
There are four basic types of dipole, corresponding
to each of the emitter and spectator partons being in
either the initial or final state. In all cases we label the partons
as $\widetilde{ik}$ emitting parton $k$ with respect to the spectator
$j$.

\subsection{Initial-initial}

We consider the case of an emitter $i$ and a spectator $j$,
both in the initial state, radiating parton $k$ into the final state.
We first define the dimensionless variable $x$ to be,
\beq \label{xdef}
x=1-\frac{s_{ik}+s_{jk}}{s_{ij}}.
\eeq
The dipole functions are then given by,
\beqn
\cD_{ii,\,qq}^{ik,j} & = & \frac{g^2 \mu^{2 \epsilon} }{x p_i \cdot p_k}
 \left( \frac{2}{1-x}-1-x \right) \nonumber \\
\cD_{ii,\,gq}^{ik,j} & = & \frac{g^2 \mu^{2 \epsilon} }{x p_i \cdot p_k}
 \left( x \right) \nonumber \\
\cD_{ii,\,qg}^{ik,j} & = & \frac{g^2 \mu^{2 \epsilon} }{x p_i \cdot p_k}
 \left( 1-2x(1-x) \right) \nonumber \\
\cD_{ii,\,gg}^{ik,j} & = & \frac{g^2 \mu^{2 \epsilon} }{x p_i \cdot p_k}
 \left( \frac{2 x}{1-x}+2 x(1-x) \right)
\label{dips:ii}
\eeqn
These are the dipole functions if we neglect spin correlations
between partons $i$ and $k$ and are thus not sufficient for the
cases $\cD_{ii,\,gq}$ and $\cD_{ii,\,gg}$.
If the lowest-order amplitude is represented by
$\cM_{\mu}$ ($\cM_{\nu}^*$), where $\mu$ ($\nu$) is the polarization
index of gluon $\widetilde{ik}$, then the required subtraction is
actually,
\beq
\left( -g^{\mu\nu} \cD^{ik,j} + q^\mu q^\nu \, \widetilde{\cD}^{ik,j}
 \right) \cM_{\mu} \cM_{\nu}^*,
\eeq 
where the additional pieces are,
\beqn
\widetilde{\cD}_{ii,\,gq}^{ik,j} & = & 
 \frac{g^2\mu^{2 \epsilon} }{x p_i \cdot p_k} \;
 \frac{1-x}{x} \frac{2 p_i \cdot p_j}{p_i \cdot p_k \, p_j \cdot p_k},
  \nonumber \\
\widetilde{\cD}_{ii,\,gg}^{ik,j} & = & 
 \widetilde{\cD}_{ii,\,gq}^{ik,j},
 \nonumber \\
 q^{\mu} & = & p_k^{\mu}-\frac{p_i \cdot p_k}{p_i \cdot p_j} p_j^{\mu}.
\eeqn
When we perform the integration, it will be over the azimuthal
average of these terms. The azimuthal average over the tensor $q^\mu q^\nu$
can only depend on the vectors $p_i$ and $p_j$ and since $q \cdot p_i=0$ it 
is equal to
\beq
\langle q^\mu q^\nu \rangle= A\Big[ -g^{\mu \nu} 
+\frac{p_i^\mu p_j^\nu +p_j^\mu p_i^\nu}{p_i \cdot p_j } \Big] +B p_i^\mu 
p_i^\nu
\eeq 
where $A=-q^2/2$. The final result is,
\beq
\langle\cD^{ik,j}\rangle = 
 \cD^{ik,j} - \frac{q^2}{2}\widetilde{\cD}^{ik,j},
\eeq
with $\widetilde{\cD}_{ii,\,qq} = \widetilde{\cD}_{ii,\,qg} =0$.

To integrate these dipoles we decompose the $n$-particle phase space in 
$d=4-2 \epsilon$ dimensions,
\beq
d\phi^{(n)}(p_a,p_b \rightarrow p_1 \ldots p_n) =
\prod_{i=1,n} \; \Big[ \frac{d^{d}p_i}{(2\pi)^{d-1}} \,\delta^+(p_i^2) \Big] \;
\;(2\pi)^d\,\delta^{(d)}(p_a+p_b-\sum_{i=1}^{n} p_i ) \;,
\eeq
into two pieces,
\beq
d\phi^{(n)}(p_i,p_j \rightarrow \ldots, k_{n-1},p_k) =  \int_0^1 dx
\; 
d\phi^{(n-1)}({\widetilde p}_i,{\widetilde p}_j \rightarrow 
 \ldots, {\widetilde k}_{n-1})
\;
\left[ dp_k(p_i,p_j, x) \right] \;,
\eeq
The first term corresponds to an $(n-1)$-particle phase-space and
the second is the one-particle dipole sub-space that we must integrate
over. Under this decomposition, the momenta of the particles are
transformed. The final state momenta all undergo a Lorentz transformation
(the details, which are given in ref.~\cite{CS} erratum, 
are unimportant here), 
The initial momenta are modified with $p_i \to
\widetilde{p}_i = x p_i$ and $p_j = \widetilde{p}_j$ is unchanged.
The emitted parton phase space is given by,
\beq \label{dipphase}
\left[ dp_k(p_i,p_j, x) \right] =
\frac{d^{d}p_k}{(2\pi)^{d-1}} \,\delta^+(p_k^2) \;
\Theta(x) \Theta(1-x) \;
\eeq
Eq.~(\ref{dipphase}) can be rewritten as,
\beqn
\label{ii:PSconv}
\left[ dp_k(p_i,p_j, x) \right]
&=& \frac{(2p_i p_j)^{1-\epsilon}}{16\pi^2}
\;\frac{d\Omega_{d-3}}{(2\pi)^{1-2\epsilon}}
\;dv \;dx
\;\Theta(x(1-x))
\;\Theta(v) \;\Theta\!\left(1-\frac{v}{1-x} \right)
\nonumber \\
&\times&
\;\left( 1-x\right)^{-2\epsilon} 
\;\left[\frac{v}{1-x}
\left(1-\frac{v}{1-x}\right) \right]^{-\epsilon} \;,
\eeqn
where $x$ is defined in Eq.~(\ref{xdef}),
$v=p_i p_k/p_i p_j$ and
$d\Omega_{d-3}$ is an element of solid angle in the directions
perpendicular to the plane defined by $p_i$ and $p_j$.
\beq
\int \frac{d\Omega_{d-3}}{(2\pi)^{1-2\ep}} = 
 \frac{(4 \pi)^\ep}{\Gamma(1-\ep)}=c_\Gamma \;\;.
\eeq
Isolating the common factor $g^2/p_i \cdot p_k$ from the dipole terms
and re-writing all the momenta in terms of the transformed ones,
we can further express this as,
\beq
\frac{g^2}{p_i \cdot p_k} \,
 [dp_k(p_i,p_j,x)] =
 \frac{d\Omega_{d-3}}{(2\pi)^{1-2\epsilon}} \,
\frac{\alpha_S}{2\pi} \left( \frac{x}{2 \widetilde{p}_i \cdot
 \widetilde{p}_j} \right)^\epsilon (1-x)^{-2\epsilon} \,
 \frac{dv}{v} \, \left(\frac{v}{1-x}\right)^{-\epsilon}
  \left(1-\frac{v}{1-x}\right)^{-\epsilon}.
\eeq
This is now in a form where we can integrate the dipoles in
Eq.~(\ref{dips:ii}). We perform the integrals 
and then expand in powers of
$\epsilon$, discarding terms of $\cal{O}(\epsilon)$. In this way we
find,
\beqn \label{rawresult}
\lefteqn{
\int \, \left[ dp_k(p_i,p_j, x) \right] \left[ x
 \langle \cD_{ii}^{ik,j} \rangle \right]
 = \left( \frac{\alpha_S}{2\pi} \right) 
 \left( \frac{x \mu^2 }
{2 \widetilde{p}_i \cdot \widetilde{p}_j } \right)^\epsilon \times c_\Gamma 
\Biggl[ }
 \nonumber \\
 && - \frac{1}{\epsilon} \, p_{qq}(x)
 +\delta(1-x)\left(\frac{1}{\epsilon^2}+\frac{3}{2\epsilon}
 -\frac{\pi^2}{6}\right) + \frac{4\ln(1-x)}{(1-x)_+}
 -2(1+x)\ln(1-x)\Biggr],
\label{qqint}
\eeqn
where we have expressed the result in terms of the splitting function,
\beq
p_{qq}(x) = \frac{2}{(1-x)_+}-1-x+\frac{3}{2}\delta(1-x).
\eeq
Before proceeding further we must perform mass factorization to get
into the $\overline{MS}$ scheme.
In $d=4-2\epsilon$ dimensions, the splitting functions contain 
additional terms
\begin{equation}
 p_{ij}(x,\epsilon)=p_{ij}(x) + \epsilon \; p_{ij}^{(\epsilon)}(x)
\end{equation}
We have the following expressions for the $p_{ij}$,
\beqn
p_{qq}(x) &=& \frac{2}{(1-x)}_+ -1 -x + \frac{3}{2} \delta(1-x)
+\epsilon \Big(-(1-x)+\frac{1}{2} \, \delta(1-x) \Big)\;, \nonumber \\
p_{gq}(x) &=& \frac{1+(1-x)^2}{x}-\epsilon \; x \;, \nonumber \\
p_{qg}(x) &=& x^2 + (1-x)^2 -\epsilon \; 2x (1-x) \;, \nonumber \\
p_{gg}^(x) &=& \frac{2}{(1-x)}_+ + \frac{2}{x} -4 +2 x (1-x) 
+ \delta(1-x) \left( \frac{11-2 n_f/N}{6} \right) \;.
\eeqn
In order to arrive at the $\overline{MS}$ scheme, starting from results 
in the four-dimensional helicity scheme given by Eq.~(\ref{rawresult}),
we must subtract the counterterm
\beq
\left( \frac{\alpha_S}{2\pi} \right) \left[
 -\frac{1}{\epsilon}p_{qq}(x) + p_{qq}^{(\epsilon)}(x) \right] \;.
\label{counterterm}
\eeq
After performing this subtraction we have,
\beqn
&&
\left( \frac{\alpha_S}{2\pi} \right) \left\{
\left( \frac{x \mu^2 }
{ 2 \widetilde{p}_i \cdot \widetilde{p}_j } \right)^\epsilon
\left[-\frac{1}{\epsilon} \, p_{qq}(x) + \ldots \right]
-\left[-\frac{1}{\epsilon} \, p_{qq}(x) + p_{qq}^{(\epsilon)}(x) 
 \right] \right\}
\nonumber \\
&=&
\left( \frac{\alpha_S}{2\pi} \right) \left\{
\left( \frac{x \mu^2 }{2 \widetilde{p}_i \cdot \widetilde{p}_j } 
\right)^\epsilon \biggl[ \, \ldots \, \biggr]
-(\ln x-L) p_{qq}(x) - p_{qq}^{(\epsilon)}(x)\right\},
\eeqn
where $\left[ \ldots \right]$ represents all the non-$p_{qq}$ terms
in Eq.~(\ref{qqint}). We have used the shorthand
$L \equiv \ln(2 \widetilde{p}_i \cdot \widetilde{p}_j / \mu^2)$.
With the above mass factorization procedure
understood, we now introduce the notation,
\beq
\int \, \left[ dp_k(p_i,p_j, x) \right] \left[ x
 \langle \cD_{ii}^{ik,j} \rangle \right]
 \equiv \left( \frac{\alpha_S}{2\pi} \right) \, c_\Gamma
 \; \left[
 \cV_{ii}^{end}\delta(1-x) + \cV_{ii}^{reg} + \cV_{ii}^{plus} \right],
\label{ii:Vdef}
\eeq
where we have split the result into an end-point contribution
proportional to $\delta(1-x)$, regular terms and a term
containing `plus'-distributions. In the $q \to q$ case we have,
\beqn
\cV_{ii,\,qq}^{end} & = & \frac{1}{\epsilon^2}+\frac{1}{\epsilon}
 \left( \frac{3}{2}-L \right) + \frac{L^2}{2} - \frac{1}{2}
 -\frac{\pi^2}{6},
 \nonumber \\
\cV_{ii,\,qq}^{reg} & = & 1-x-\frac{1+x^2}{1-x}\ln x
 -(1+x)\left( L + 2\ln(1-x) \right),
 \nonumber \\
\cV_{ii,\,qq}^{plus} & = & \frac{2 L }{(1-x)_+} 
 +4\Bigg[\frac{\ln(1-x)}{1-x} \Bigg]_+.
\eeqn
The expressions for the other parton splittings,
derived in the same manner are,
\beqn
\cV_{ii,\,qg}^{end} & = & \cV_{ii,\,qg}^{plus} \, = \, 0,
 \nonumber \\
\cV_{ii,\,qg}^{reg} & = & (1-2x(1-x)) (L-\ln x+2\ln(1-x))+2x(1-x),
 \nonumber \\
\cV_{ii,\,gq}^{end} & = & \cV_{ii,\,gq}^{plus} \, = \, 0,
 \nonumber \\
\cV_{ii,\,gq}^{reg} & = & \left( \frac{1+(1-x)^2}{x} \right)
 (L-\ln x+2\ln(1-x))+x,
 \nonumber \\
\cV_{ii,\,gg}^{end} & = & 
\frac{1}{\epsilon^2} - \frac{L}{\epsilon}
 + \frac{L^2}{2} - \frac{\pi^2}{6}
 +\frac{1}{6\epsilon} \left( 11 - \frac{2n_f}{N} \right),
 \nonumber \\
\cV_{ii,\,gg}^{reg} & = & 2(L-\ln x + 2\ln(1-x)) \left(
 \frac{1-x}{x}+x(1-x)-1 \right) -\frac{2\ln x}{1-x},
 \nonumber \\
\cV_{ii,\,gg}^{plus} & = & \frac{2 L }{(1-x)_+} 
 +4\Bigg[\frac{\ln(1-x)}{1-x}\Bigg]_+.
\eeqn

\subsection{Initial-final}

We now consider the case of an initial emitter $i$ with respect to 
a final state spectator $j$. We first define the dimensionless variables $x$ 
and $u$ to be,
\beq \label{xudef}
x=1-\frac{s_{jk}}{s_{ik}+s_{ij}}, \qquad u=\frac{s_{ik}}{s_{ik}+s_{ij}}.
\eeq
The dipole functions are then given by,
\beqn
\cD_{if,\,qq}^{ik,j} & = & \frac{g^2 \mu^{2 \epsilon} }{x p_i \cdot p_k}
 \left( \frac{2}{1-x+u}-1-x \right), \nonumber \\
\cD_{if,\,gq}^{ik,j} & = & \frac{g^2 \mu^{2 \epsilon} }{x p_i \cdot p_k}
 \left( x \right), \nonumber \\
\cD_{if,\,qg}^{ik,j} & = & \frac{g^2 \mu^{2 \epsilon} }{x p_i \cdot p_k}
 \left( 1-2x(1-x) \right), \nonumber \\
\cD_{if,\,gg}^{ik,j} & = & \frac{g^2 \mu^{2 \epsilon} }{x p_i \cdot p_k}
 \left( \frac{2}{1-x+u}-2+2 x(1-x) \right),
\label{dips:if}
\eeqn
with the additional variables needed to account for spin correlations,
\beqn
\widetilde{\cD}_{if,\,gq}^{ik,j} & = & 
  \frac{g^2 \mu^{2 \epsilon} }{ x p_i \cdot p_k}
 \, \frac{2(1-x)}{x} \, \frac{u(1-u)}{p_j \cdot p_k},
  \nonumber \\
\widetilde{\cD}_{if,\,gg}^{ik,j} & = & 
 \widetilde{\cD}_{if,\,gq}^{ik,j},
 \nonumber \\
 q^{\mu} & = & \frac{p_k^{\mu}}{u}-\frac{p_j^\mu}{1-u}.
\eeqn

The phase-space convolution becomes
\beq
\label{psconva}
d\phi^{(n)}(p_i,p_b\rightarrow \ldots, p_j,p_k) =  \int_0^1 dx \; 
d\phi^{(n-1)}(\widetilde{p}_i, p_b \rightarrow \ldots, {\widetilde p}_{j}) \;
\left[ dp_k({\widetilde p}_{j};p_i, x) \right] \;,
\eeq
where the transformed momenta $\widetilde{p}_i$ and $\widetilde{p}_j$
are defined in terms of the original momenta by,
\beq
\widetilde{p}_i = xp_i , \qquad
\widetilde{p}_j = p_j+p_k+(1-x)p_i,
\label{if:ptilde}
\eeq
and $\widetilde{p}_i^2 = \widetilde{p}_j^2 = 0$. The dipole phase space
is,
\beq
\label{dpix}
\left[ dp_k({\widetilde p}_{j}; p_i, x) \right] =
\frac{d^{d}p_k}{(2\pi)^{d-1}} \,\delta^+(p_k^2) \;
\Theta(x) \Theta(1-x) \; \frac{1}{1-u} \;\;.
\eeq
Using the kinematic variables in Eq.~(\ref{xudef}), the phase space
in Eq.~(\ref{dpix}) can be written as follows
\beqn
\label{exdpix}
\left[ dp_k({\widetilde p}_{j};p_i, x) \right]
&=& \frac{(2{\widetilde p}_{j} p_i)^{1-\ep}}{16\pi^2}
\;\frac{d\Omega_{d-3}}{(2\pi)^{1-2\ep}}
\;du \;dx
\;\Theta(u(1-u))
\;\Theta(x(1-x)) \nonumber \\
&\cdot&
\;\left(u(1-u) \right)^{-\ep} \;\left( 1-x \right)^{-\ep} \;
\eeqn
where $d\Omega_{d-3}$ is an element of solid angle perpendicular to
${\widetilde p}_{k}$ and $p_1$.

We manipulate further as before to yield,
\beq
\frac{g^2}{p_i \cdot p_k} \,
\left[ dp_k({\widetilde p}_{j}; p_i, x) \right] =
 \frac{d\Omega_{d-3}}{(2\pi)^{1-2\epsilon}} \,
\frac{\alpha_S}{2\pi} \left( \frac{x}{2 \widetilde{p}_i \cdot
 \widetilde{p}_j} \right)^\epsilon (1-x)^{-\epsilon}
 du \, u^{-1-\epsilon} (1-u)^{-\epsilon}.
\eeq

The integration of the dipoles with this phase space and the mass
factorization procedure are handled in the same way as the
initial-initial case detailed above. The only challenging integral is given by
\begin{eqnarray}
I_{if}&=& \int_0^1 du \; u^{-1-\epsilon} (1-u)^{-\epsilon} 
\frac{1}{(1-x+u)} \nonumber \\
&=&-\frac{1}{\epsilon} \frac{\Gamma^2(1-\epsilon)}{\Gamma(1- 2\epsilon)}
\frac{(2-x)^{\epsilon}}{(1-x)^{1+\epsilon}}
\; F(-\epsilon,-2\epsilon ; 1-2\epsilon; \frac{1}{2-x})
\end{eqnarray}
where $F(-\epsilon,-2\epsilon ; 1-2\epsilon; \frac{1}{2-x})
=1+{\cal O}(\epsilon^2(1-x)^0)$.
Making the separation of
the integrated dipoles into different types of contribution yields,
(in a notation similar to Eq.~(\ref{ii:Vdef})).
\beqn
\cV_{if,\,qq}^{end} & = & \frac{1}{\epsilon^2}+\frac{1}{\epsilon}
 \left( \frac{3}{2}-L \right) + \frac{L^2}{2} - \frac{1}{2}
 +\frac{\pi^2}{6},
 \nonumber \\
\cV_{if,\,qq}^{reg} & = & 1-x-\frac{2\ln(2-x)}{1-x}
 -(1+x)\left( L-\ln x + \ln(1-x) \right) - \frac{2\ln x}{1-x},
 \nonumber \\
\cV_{if,\,qq}^{plus} & = & 
\frac{2L}{(1-x)_+} +4 \Bigg[\frac{\ln(1-x)}{1-x}\Bigg]_+.
 \nonumber \\
\cV_{if,\,qg}^{end} & = & \cV_{if,\,qg}^{plus} \, = \, 0,
 \nonumber \\
\cV_{if,\,qg}^{reg} & = & (1-2x(1-x)) (L-\ln x+\ln(1-x))+2x(1-x),
 \nonumber \\
\cV_{if,\,gq}^{end} & = & \cV_{if,\,gq}^{plus} \, = \, 0,
 \nonumber \\
\cV_{if,\,gq}^{reg} & = & \left( \frac{1+(1-x)^2}{x} \right)
 (L-\ln x+\ln(1-x))+x,
 \nonumber \\
\cV_{if,\,gg}^{end} & = & \frac{1}{\epsilon^2}-\frac{L}{\epsilon}
 + \frac{L^2}{2} +\frac{\pi^2}{6}
 +\frac{1}{6\epsilon} \left( 11 - \frac{2n_f}{N} \right),
 \nonumber \\
\cV_{if,\,gg}^{reg} & = & 2(L-\ln x + \ln(1-x)) \left(
 \frac{1-x}{x}+x(1-x)-1 \right) -\frac{2\ln(2-x)}{1-x}
 -\frac{2\ln x}{1-x},
 \nonumber \\
\cV_{if,\,gg}^{plus} & = & \frac{2L}{(1-x)_+} 
+4 \Bigg[ \frac{\ln(1-x)}{1-x} \Bigg]_+ .
\eeqn

\subsection{Final-initial}

We now consider a final state emitter $i$, with respect to an initial 
state spectator $j$.
We first define the dimensionless variables $x$ and $z$ to be,
\beq
x=1-\frac{s_{ik}}{s_{ij}+s_{jk}}, \qquad z=\frac{s_{ij}}{s_{ij}+s_{jk}}.
\eeq
The dipole functions are then given by,
\beqn
\cD_{fi,\,qq}^{ik,j} & = & \frac{g^2 \mu^{2 \epsilon} }{x p_i \cdot p_k}
 \left( \frac{2}{1-x+1-z}-1-z \right), \nonumber \\
\cD_{fi,\,gq}^{ik,j} & = & \frac{g^2 \mu^{2 \epsilon} }{x p_i \cdot p_k},
 \nonumber \\
\cD_{fi,\,gg}^{ik,j} & = & \frac{g^2 \mu^{2 \epsilon} }{x p_i \cdot p_k}
 \left( \frac{2}{1-x+1-z}+\frac{2}{1-x+z}-4 \right),
\eeqn
There is no dipole $\cD_{fi,\,qg}$ since the singularities are 
already accounted for by $\cD_{fi,\,qq}$. 
The dipole pieces associated with spin
correlations are given by,
\beqn
\widetilde{\cD}_{fi,\,gq}^{ik,j} & = & 
\frac{g^2 \mu^{2 \epsilon}}{x p_i \cdot p_k}
 \, \frac{-2}{p_i \cdot p_k},
  \nonumber \\
\widetilde{\cD}_{fi,\,gg}^{ik,j} & = & 
  - \widetilde{\cD}_{fi,\,gq}^{ik,j},
 \nonumber \\
 q^{\mu} & = & z p_i^{\mu}-(1-z)p_k^\mu.
\eeqn

The phase-space has the following convolution structure,
\beq 
d\phi^{(n)}(p_a,p_j \rightarrow \ldots, p_i,p_k) =  
\int_0^1 dx \; 
d\phi^{(n-1)}(p_a,x p_j,\rightarrow \ldots, \widetilde{p}_{ik})\;
\left[ dp_k({\widetilde p}_{i} ;p_j, x) \right] \;,
\eeq
where
\beq \label{px}
\left[ dp_k({\widetilde p}_{i}; p_j, x) \right] =
\frac{d^{d}p_k}{(2\pi)^{d-1}} \,\delta^+(p_k^2) \;
\Theta(x) \Theta(1-x) \; \frac{1}{1-z} \;\;,
\eeq
and the transformed momenta are given by,
\begin{eqnarray}
\widetilde{p}_{i} & = & p_i+p_k+(1-x)p_j\nonumber \\
\widetilde{p}_{j} & = & x p_j.
\end{eqnarray}
Equation~(\ref{px}) can be written out more explicitly as,
\beqn
\label{expx}
\left[ dp_k({\widetilde p}_{i};p_j, x) \right]
&=& \frac{(2{\widetilde p}_{i}p_j)^{1-\ep}}{16\pi^2}
\;\frac{d\Omega_{d-3}}{(2\pi)^{1-2\ep}}
\;dz \;dx
\;\Theta(z(1-z))
\;\Theta(x(1-x)) \nonumber \\
&\times&
\;\left(z(1-z) \right)^{-\ep}
\;\left( 1-x \right)^{-\ep} \;,
\eeqn
where $d\Omega_{d-3}$ is an element of solid angle perpendicular to
the plane defined by ${\widetilde p}_{i}$ and $p_j$.

The form needed to integrate the dipoles is then,
\beq
\frac{g^2}{p_i \cdot p_k} \,
 [dp_k(\widetilde{p}_{ik};p_j,x)] =
 \frac{d\Omega_{d-3}}{(2\pi)^{1-2\epsilon}} \,
\frac{\alpha_S}{2\pi} \left( \frac{x}{2 \widetilde{p}_i \cdot
 \widetilde{p}_j} \right)^\epsilon (1-x)^{-1-\epsilon}
 dz \, z^{-\epsilon} (1-z)^{-\epsilon},
\eeq
and the integrations over $z$ can now be performed. The only difficult
integral is
\begin{eqnarray}
I_{fi}&=& \int_0^1 dz \; z^{-\epsilon} (1-z)^{-\epsilon} \frac{1}{(1-x+z)}
=-\frac{1}{\epsilon} \frac{\Gamma^2(1-\epsilon)}{\Gamma(1-2 \epsilon)}
\nonumber \\
&&
 \times \Bigg[ (2-x)^{-2 \epsilon} 
F(\epsilon,2\epsilon;1+\epsilon;\frac{1-x}{2-x})
  - \Big((1-x)(2-x)\Big)^{- \epsilon}
\frac{\Gamma(1-2 \epsilon) \Gamma(1+\epsilon)}{\Gamma(1-\epsilon)} \Bigg]
\end{eqnarray}
where $F(\epsilon,2\epsilon;1+\epsilon;z)=1+O(\epsilon^2 z)$.
With definitions
analogous to Eq.~(\ref{ii:Vdef}) we find,
\beqn
\cV_{fi,\,qq}^{end} & = & 
 \frac{1}{\epsilon^2} + \frac{1}{\epsilon} \left( \frac{3}{2} - L
 \right) + \frac{L^2}{2} - \frac{3L}{2} + 3 - \frac{\pi^2}{2},
 \nonumber \\
\cV_{fi,\,qq}^{reg} & = & \frac{2\ln(2-x)}{1-x},
 \nonumber \\
\cV_{fi,\,qq}^{plus} & = & -\Bigg[\frac{2\ln(1-x)}{(1-x)}\Bigg]_+
 -\frac{3}{2(1-x)_+}.
 \nonumber \\
\cV_{fi,\,gq}^{end} & = & - \frac{2}{3\epsilon} - \frac{13}{9} +\frac{2 L }{3},
 \nonumber \\
\cV_{fi,\,gq}^{reg} & = & 0,
 \nonumber \\
\cV_{fi,\,gq}^{plus} & = & \frac{2}{3} \, \frac{1}{(1-x)_+},
 \nonumber \\
\cV_{fi,\,gg}^{end} & = & \frac{2}{\epsilon^2}
 +\frac{1}{\epsilon}(\frac{11}{3}-2L)
-\frac{11 L}{3} + L^2 + \frac{67}{9} - \pi^2,
 \nonumber \\
\cV_{fi,\,gg}^{reg} & = & \frac{4\ln(2-x)}{1-x},
 \nonumber \\
\cV_{fi,\,gg}^{plus} & = & -\Bigg[\frac{4\ln(1-x)}{1-x} \Bigg]_+
 -\frac{11}{3(1-x)_+}.
\eeqn

\subsection{Final-final}

The remaining dipole to consider is one in which we have a final
state emitter $i$ with respect to a final state spectator $j$.
In this case we define the dimensionless variables $y$ and $z$,
\beq
y=\frac{s_{ik}}{s_{ik}+s_{ij}+s_{jk}}, \qquad
z=\frac{s_{ij}}{s_{ij}+s_{jk}}.
\eeq
The dipole functions are then given by,
\beqn
\cD_{ff,\,qq}^{ik,j} & = & \frac{g^2 \mu^{2 \epsilon} }{ p_i \cdot p_k}
 \left( \frac{2}{1-z(1-y)}-1-z \right), \nonumber \\
\cD_{ff,\,gq}^{ik,j} & = & \frac{g^2 \mu^{2 \epsilon} }{ p_i \cdot p_k}
 , \nonumber \\
\cD_{ff,\,gg}^{ik,j} & = & \frac{g^2 \mu^{2 \epsilon} }{p_i \cdot p_k}
 \left( \frac{2}{1-z(1-y)}+\frac{2}{1-(1-z)(1-y)}-4 \right),
\eeqn
with the auxiliary information,
\beqn
\widetilde{\cD}_{ff,\,gq}^{ik,j} & = & \frac{g^2 \mu^{2 \epsilon} }{p_i 
\cdot p_k}
 \, \frac{-2 }{p_i \cdot p_k},
  \nonumber \\
\widetilde{\cD}_{ff,\,gg}^{ik,j} & = & -
 \widetilde{\cD}_{ff,\,gq}^{ik,j},
 \nonumber \\
 q^{\mu} & = & z p_i^{\mu}-(1-z)p_k^\mu.
\eeqn
In terms of the momenta ${\widetilde p}_{i}, \,
{\widetilde p}_j$ and $p_k$, this phase-space
contribution takes the factorized form:
\beqn \label{psfac}
d\phi^{(n)}(p_a,p+b \rightarrow \ldots, p_i,p_k,p_j) = 
d\phi^{(n-1)}(p_a,p+b \rightarrow \ldots, 
{\widetilde p}_{i},{\widetilde p}_j)
\left[ dp_k({\widetilde p}_{i},{\widetilde p}_j) \right] \;\;,
\eeqn
where
\beqn
\label{dpi}
\left[ dp_k({\widetilde p}_{i},{\widetilde p}_j) \right]
= \frac{d^{d}p_k}{(2\pi)^{d-1}} \,\delta^+(p_k^2) \;
{\cal J}(p_k;{\widetilde p}_{i},{\widetilde p}_j) \;,
\eeqn
and the Jacobian factor is
\beq
\label{Jac}
{\cal J}(p_k;{\widetilde p}_{i},{\widetilde p}_j)
= \Theta(1- z) \,\Theta(1-y) \;
\frac{(1-y)^{d-3}}{1- z} \;.
\eeq
In terms of the kinematic variables defined earlier, we have
\beqn
\label{exdpi}
\left[ dp_k({\widetilde p}_{i},{\widetilde p}_j) \right]
&=& \frac{(2{\widetilde p}_{i}{\widetilde p}_j)^{1-\ep}}{16\pi^2}
\;\frac{d\Omega_{d-3}}{(2\pi)^{1-2\ep}}
\;dz \;dy
\;\Theta(z(1-z))
\;\Theta(y(1-y)) \nonumber \\
&\times &
\;\left(z(1-z) \right)^{-\ep}
\;\left( 1-y \right)^{1-2\ep} y^{-\ep} \;,
\eeqn
where $d\Omega_{d-3}$ is an element of solid angle perpendicular to
${\widetilde p}_{ij}$ and ${\widetilde p}_k$.
In this case the momenta transform as,
\beq
\widetilde{p}_i=p_i+p_k-\left(\frac{y}{1-y}\right) \, p_j, \qquad
\widetilde{p}_j=\frac{1}{(1-y)} \, p_j,
\eeq
and the phase-space factorizes to yield the relevant measure,
\beq
\frac{g^2}{p_i \cdot p_k} \,
 [dp_k(\widetilde{p}_i,\widetilde{p}_j,x)] =
 \frac{d\Omega_{d-3}}{(2\pi)^{1-2\epsilon}} \,
\frac{\alpha_S}{2\pi} \left( \frac{1}{2 \widetilde{p}_i \cdot
 \widetilde{p}_j} \right)^\epsilon dz \, z^{-\epsilon} (1-z)^{-\epsilon}
 \, dy \, y^{-1-\epsilon} (1-y)^{1-2\epsilon}.
\eeq

The soft integral is then given by
\begin{eqnarray}
I_{ff}&=& \int_0^1 dz \; z^{-\epsilon} (1-z)^{-\epsilon}
\int_0^1 dy \; y^{-1-\epsilon} (1-y)^{1-2\epsilon} \frac{1}{1-z(1-y)}
\nonumber \\
&=& 
\frac{1}{2 \epsilon^2} \frac{\Gamma^3(1-\epsilon)}{\Gamma(1-3 \epsilon)}
\end{eqnarray}
In this case, there is no need to separate the integrated forms
of these dipoles and with the simplified notation,
\beq
\int \, [dp_k(\widetilde{p}_i,\widetilde{p}_j,x)] \left[
 \langle \cD_{ff}^{ik,j} \rangle \right]
 \equiv \left( \frac{\alpha_S}{2\pi} \right) \, c_\Gamma \; \cV_{ff},
\label{ff:Vdef}
\eeq
we find,
\beqn
\cV_{ff,\,qq} & = & 
 \frac{1}{\epsilon^2} + \frac{1}{\epsilon} \left( \frac{3}{2} - L
 \right) + \frac{L^2}{2} - \frac{3L}{2} + \frac{9}{2} - \frac{\pi^2}{2},
 \nonumber \\
\cV_{ff,\,gq} & = & - \frac{2}{3\epsilon} - \frac{19}{9}+\frac{2L}{3},
 \nonumber \\
\cV_{ff,\,gg} & = & \frac{2}{\epsilon^2} + \frac{1}{\epsilon} \left(
 \frac{11}{3} - 2L \right) + L^2 - \frac{11L}{3}
 + \frac{100}{9} - \pi^2.
\eeqn

\end{document}